# On the production of flat electron bunches for laser wake field acceleration

M. Kando, Y. Fukuda, H. Kotaki, J. Koga, S. V. Bulanov[†], T. Tajima

*Kansai Photon Science Institute, Japan Atomic Energy Agency, Kyoto 619-0215, Japan*

A. Chao, R. Pitthan

*Stanford Linear Accelerator Center, Menlo Park, California 94025, USA*

K.-P. Schuler

*DESY, Deutsches Elektronen Synchrotron, 22603 Hamburg, Germany*

A. G. Zhidkov[†], K. Nemoto

*Central Research Institute of Electric Power Industry, Yokosuka, Kanagawa 240-0196, Japan*

**Abstract**

We suggest a novel method for injection of electrons into the acceleration phase of particle accelerators, producing low emittance beams appropriate even for the demanding high energy Linear Collider specifications. In this paper we work out the injection into the acceleration phase of the wake field in a plasma behind a high intensity laser pulse, taking advantage of the laser polarization and focusing. With the aid of catastrophe theory we categorize the injection dynamics. The scheme uses the structurally stable regime of transverse wake wave breaking, when electron trajectory self-intersection leads to the formation of a flat electron bunch. As shown in three-dimensional particle-in-cell simulations of the interaction of a laser pulse in a line-focus with an underdense plasma, the electrons, injected via the transverse wake wave breaking and accelerated by the wake wave, perform betatron oscillations with different amplitudes and frequencies along the two transverse coordinates. The polarization and focusing geometry lead to a way to produce relativistic electron bunches with asymmetric emittance (flat beam). An approach for generating flat laser accelerated ion beams is briefly discussed.

---

[†] Also at A. M. Prokhorov Institute of General Physics of Russian Academy of Sciences, Moscow 119991, Russia.

# 1. Introduction

Electron accelerators with energies of many GeV and low emittance are needed for coherent light sources and linear colliders. The laser acceleration of charged particles provides a promising approach toward such development in a compact way, avoiding some of the complications which arise for the additional requirements of asymmetric emittance for linear colliders, as outlined below. In the Laser Wake Field Accelerator (LWFA), concept electrons are accelerated by the longitudinal electric field created in an underdense plasma by a high intensity short pulse laser [1]. Electrons injected by conventional accelerators, or self-injected by nonlinear wakewave breaking (see for details a review article [2] and literature therein) achieve energies substantially higher than the initial injection energies. Although the understanding and production of high intensity ( $\approx$nC) and low emittance ($\approx$ 2-3 mm mrad) electron beams via laser plasma interaction has made rapid progress in recent years [3,4-7], applications to coherent light sources and linear colliders still demand further advances. A symmetric emittance of about 1mm-mrad is needed for coherent light sources to reach X-ray wavelengths of 1 Ångstrom [8]. Linear electron-positron colliders need asymmetric emittances and polarized electrons, with the smaller vertical emittance required to be of the order of 0.1 mm-mrad. The asymmetric emittances (flat beam[1]) are needed to reduce the beam induced synchrotron radiation (beamstrahlung) in the interaction (see below), and electron polarization is required because the effective luminosity can be up to 2 orders of magnitude larger (depending on the process), as shown by the SLD experiment at the Stanford Linear Collider (SLC) [9]. Additionally, positron beam polarization is desirable because certain processes can be measured with higher signal-to-noise ratio due to the additional positron polarization, and it enables the measurement of transverse cross sections [10]. The capability to selectively suppress unwanted background processes is especially desired in search for new physics.

---

[1] This paper addresses both laser-plasma and accelerator physics issues. One frequent confusion in the nomenclature is the use of the term "flat beam". In accelerator physics flat beam means an asymmetric spot size (created with an asymmetric emittance). In laser physics a flat beam is one with a large focal length, and the means to achieve an asymmetric spot size is a "line focus".

With the near-term aim of producing 1 GeV range laser accelerated electron bunches, a more quantitative evaluation of differing requirements of major applications using GeV electrons is in order. These requirements include, in order of rising beam quality, 1) fixed target – electron beam interactions, 2) synchrotron and coherent light sources and 3) the colliding electron (positron) beam configurations:

1. For fixed target applications the electron beam emittance is less important because the luminosity is determined by the beam charge and the target thickness [11]. In this application the electron polarization can play a role and enhance the effective luminosity.

2. In case of the GeV electron beam usage for coherent light sources, e. g. for linac light sources, low emittance is important, because radiation from a high emittance beam fails to be coherent. This application requires small emittances, of the order of 1 mm-mrad for 1Å wavelength, but a small spot size is not necessary.

3. For colliding electron beams, the emittance determines the minimum achievable spot size, i.e. it is directly related to the maximum luminosity. Depending on the electron energy, one needs a spot size approximately equal to 400 nm for 100 GeV electrons, and 5 nm (small dimension) in the case of the 1 TeV collider [12]. Required for colliders in addition are the electron beam polarization (80% or more) and flat beams with an aspect ratio of about 100, in order to reduce beamstrahlung losses and electron-positron pair production [9,13].

The conditions to have beams of high electric charge and low emittance are contradictory, because space charge effects make the transverse emittance grow [14]. The emittance is calculated as a product of the spot size, $\sigma_x$, and the divergence, $\sigma_x'$, both determined at a beam waist (or a pin hole), i.e. the transverse emittance is $\varepsilon = \sigma_x \sigma_x'$. In addition, the normalized emittance, defined as $\varepsilon_N = \gamma_e \varepsilon$, is an adiabatic invariant under beam acceleration, where $\gamma_e$ is the electron relativistic gamma factor. The values of emittance quoted earlier refer to the normalized emittances.

If we consider LWFA produced electrons, we see femtosecond range electron bunches accelerated to hundreds of MeV, and driven out of the plasma [3]. The energy spectrum is a quasi-mono-energetic form [4-7,15], which is mainly related to the fact that fast electrons reach the maximum energy and are localized at the top of the separatrix in the $x,p_x$ phase plane [16-18]. We remark two properties of LWFA ejected relativistic electrons: a) the product of the bunch length and the energy spread, the longitudinal emittance, is comparable to conventional RF sources (in the range of MeV-ps), while the very short bunch length is achieved even without bunch compression, and b) the micron-size transverse spot of the initial electron bunch corresponds to the laser spot size, which in turn may lead to a small transverse emittance. At present the emittance requirements, including the asymmetry, are met with the use of expensive damping rings, which often yield long bunches and therefore require subsequent bunch compressions. There is another R&D effort underway to produce asymmetric beams using RF guns [19].

As recognized in Ref. [16], the electron injection into the acceleration phase of a wakefield takes place due to nonlinear Langmuir wave wave-breaking. This breaking of the Langmuir wave, known since J. Dawson's publication in Ref. [20], has been studied theoretically and experimentally (see Ref. [2] and literature therein). It is important to notice that the realization of resilience against wave-breaking in the relativistic regime in the longitudinal direction led to the original LWFA suggestion [1]. In the non-one-dimensional case and in the case of an inhomogeneous plasma density the wake wave-breaking acquires features that allow manipulation of the injected electron bunch parameters. For example, properties of the transverse wake wave-breaking [21] have been used in Refs. [4,15,22] in order to explain nonlinear wake evolution and electron acceleration in homogeneous and inhomogeneous plasmas. In addition, longitudinal breaking was invoked to describe electron self-injection in homogeneous plasmas [23,24], and the controllable electron injection regimes in plasmas with a tailored density profile [25].

In the present paper we shall formulate an approach for producing asymmetric emittance electron bunches by using asymmetric laser pulse focusing for the laser wakefield acceleration, when the transverse wake breaking leads to the formation of an electron bunch elongated along the transverse direction.

## 2. Laser Line Focus

In present day chirped pulse amplification (CPA) based laser systems, laser pulses have a Gaussian (TEM00)-like spatial profile. There are several schemes to achieve a line focus, which has an asymmetric transverse spatial profile. One is using astigmatism on the sagittal plane of a spherical mirror. This technique was applied to produce transient X-ray lasers. This scheme can generate a large aspect ratio of 300:1; however, the focusing depth is limited and a time difference for arriving at the focus arises [26]. A slightly misaligned off-axis parabolic mirror generates astigmatism, resulting in an asymmetric focal spot. A toroidal mirror can be used. Here, we propose to use a pair of reflective cylindrical mirrors. Cylindrical mirrors are easy to fabricate and cost-saving. A large amount of astigmatism of cylindrical mirrors placed off-axis can be improved when the incident angle is chosen properly. Further study is needed to examine the effect of effective pulse elongation in time using these line focus techniques in the generation of wakefields.

## 3. Structure of Transverse Wake Wave Breaking

Due to the nonlinear dependence of the Langmuir wave frequency on its amplitude, caused by the electron density redistribution along the laser pulse axis and by the relativistic dependence of the Langmuir frequency on the kinetic energy of the electrons, surfaces of constant phase in the wake wave give rise to a paraboloidal form [27]. The curvature of constant phase surfaces increases with the distance from the laser pulse until the curvature radius $R$ becomes comparable to the electron displacement $\varsigma$ in the wake, leading to electron trajectory self-intersection. This is the so-called regime of transverse wake breaking, which leads to electron injection into the acceleration phase [21]. Along the lines of Refs. [21, 28] we consider a wakefield plasma wave excited by a laser pulse of finite width. The condition of the wake excitation determines $\omega_w$ and $k_w$ which are the wake frequency and wavenumber, respectively, where $\omega_w = k_w v_g$, and $v_g$ is the group velocity of the driver laser pulse. The wake wave frequency, equal to the local value of the Langmuir frequency, depends on the transverse coordinates, $y$ and $z$. This dependence arises due to the plasma outward motion caused by the laser pulse ponderomotive pressure and by the relativistic dependence of the Langmuir frequency on the wave amplitude. The dependence on the wave amplitude is determined by the laser pulse transverse shape, which can be approximated in the vicinity of the axis as $a(y,z) \approx a_0 \left[ 1 - (y/s_y)^2 - (z/s_z)^2 \right]$, i.e. in the transverse plane it has an elliptic form. The wake frequency near the axis may be approximated by a simple form $\omega_w(y,z) \approx \omega_{w,0} + \Delta\omega_w \left[ (y/s_y)^2 + (z/s_z)^2 \right]$, where $s_y$ and $s_z$ are related to the curvature radii in the $y$ and $z$ directions, $\Delta\omega_w$ is the difference between the Langmuir frequency outside and on the axis of the wake field, $\omega_{w,0}$. From the expression of constant phase surfaces in the wake wave, $\psi_w(x,y,z,t) = \omega_w(y,z)(t - x/v_g) =$ constant, it follows that their curvature increases with the distance $l$ from the laser pulse front, i. e. the curvature radii decrease as $R_y = \omega_{w,0} s_y^2 / 2\Delta\omega_w l$ and $R_z = \omega_{w,0} s_z^2 / 2\Delta\omega_w l$ with $l = \psi_w v_g / \omega_{w,0}$.

We write the equation for the constant phase surface in the form

$$M_0(y_0, z_0) = \left( \frac{y_0^2}{2R_y} + \frac{z_0^2}{2R_z} \right) e_x + y_0 e_y + z_0 e_z, \tag{1}$$

where $(e_x, e_y, e_z)$ are the unit vectors along the x, y, and z axes. In the nonlinear wake the actual position of the constant phase surface is given by the equation

$$M(y_0, z_0) = M_0(y_0, z_0) + \varsigma\, n(y_0, z_0). \tag{2}$$

Here $\varsigma$ is the amplitude of the electron displacement and

$$n(y_0, z_0) = \frac{\partial_{y_0} M \times \partial_{z_0} M}{\left| \partial_{y_0} M \times \partial_{z_0} M \right|} \tag{3}$$

is the unit vector normal to the surface of constant phase. In the case when the surface is given by Eq. (1) the normal vector is

$$n(y_0, z_0) = \left[ 1 + \left( \frac{y_0}{R_y} \right)^2 + \left( \frac{z_0}{R_z} \right)^2 \right]^{-1/2} \left( e_x + \frac{y_0}{R_y} e_y + \frac{z_0}{R_z} e_z \right). \tag{4}$$

Writing in the components Eq. (2), we obtain

$$x = \frac{y_0^2}{2R_y} + \frac{z_0^2}{2R_z} + \frac{\varsigma}{\sqrt{1 + (y_0/R_y)^2 + (z_0/R_z)^2}}, \tag{5}$$

$$y = y_0 - \frac{\varsigma y_0}{R_y \sqrt{1 + (y_0/R_y)^2 + (z_0/R_z)^2}}, \tag{6}$$

$$z = z_0 - \frac{\varsigma z_0}{R_z \sqrt{1 + (y_0/R_y)^2 + (z_0/R_z)^2}}. \tag{7}$$

The mapping given by Eqs. (4) – (7) has a singularity when the Jacobian, $\left| \partial(y,z)/\partial(y_0,z_0) \right|$, vanishes. Assuming the curvature radius in the z direction to be larger than in the y direction,

$R_z > R_y$, in the limit of a relatively small but finite values of $y_0/R_y$ and $z_0/R_z$, we find that the position of the singularity in the $y_0, z_0$ plane is determined by the equation

$$2R_y = \varsigma\left[2 - 3(y_0/R_y)^2 - (z_0/R_z)^2\right]. \tag{8}$$

This equation has a solution if $\varsigma \geq R_y$, i. e. the displacement is larger than the radius of curvature. For $\varsigma > R_y$ the curve determined by Eq. (8) is an ellipse with the semi-axes $R_y\sqrt{2(\varsigma - R_y)/3\varsigma}$ and $R_z\sqrt{2(\varsigma - R_y)/\varsigma}$ in the $y_0$ and $z_0$ directions, respectively.

Plotting the surface in the $x, y, z$ space in which $\varsigma = 0$ is the paraboloid, $x = y_0^2/2R_y + z_0^2/2R_z$, $y = y_0$, $z = z_0$. With Eqs. (5)-(7) we obtain the constant phase surfaces for cases $\varsigma \neq 0$ presented in Figs. 1 and 2.

In both cases, in Figs. 1 and 2, the projections of the constant phase surface onto the $x, y$ plane have the form of a "swallow tail". This corresponds to one of the forms of fundamental catastrophes (see. [29]).

When the displacement value is in between the curvature radii, i.e. $R_y < \varsigma < R_z$ as in the case shown in Fig. 1, the singularity in the $(y, z)$ plane is elongated along the z axis. On the other hand, if $R_y < R_z < \varsigma$, the singularity in the $(y, z)$ plane is elongated along the y axis, as shown in Fig. 2 c. These types of singularities are typical, or, in other words, are structurally stable. In the 3D configurations they correspond to the the transverse wave breaking with the injection into the acceleration phase of flat electron bunches.

In the axially symmetric geometry when the curvature radii along the $y$ and $z$ directions equal to each other, $R_y = R_z$, the injected electron bunch also has axial symmetry. However, this configuration is not structurally stable and small perturbations of a general type transform it into a structurally stable configuration with non-axially-symmetric electron bunch. We point out that in the case of the gradual increase of the wake wave curvature the first breaking occurs under the condition when the electron displacement becomes greater than the minimal curvature radius, e.g. when $R_y < \varsigma < R_z$, and the injected electron bunch is elongated along the minimal curvature direction as in Fig. 1.

After the electron trajectory self-intersection has occurred, the injected electrons are accelerated and they perform betatron oscillations in the transverse direction. This stage of the electron bunch evolution is discussed below in Sections 4 and 5, where we present the results of the particle-in-cell (PIC) simulations and the analytical theory of betatron oscillations when the effects of the space charge are taken into account.

## 4. Results of Simulations of Transverse Wake Wave Breaking and Electron Bunch Injection

The paraboloidal structures of the wake plasma wave have been seen routinely in the three dimensional particle-in-cell simulations of high intensity laser pulse propagation in underdense plasmas, e. g. see Refs. [30, 31] where the laser pulse frequency upshifting has been discussed in counter- and co-propagating two pulse interaction configurations. In Refs. [6, 31] 3D PIC simulations distinctly show the "swallow tail" structure in the electron density distribution formed in the nonlinear wake wave. In this Section we present 3D PIC simulation results for the electron bunch injection, which clearly demonstrate the elongated electron bunch generation during the transverse wake wave breaking. We use the electromagnetic relativistic PIC computer code FPLaser3D [24], which exploits the moving window technique and the density decomposition scheme of the current assignment with bell-shaped quasiparticles [32]; this weighting scheme reduces significantly unphysical numerical effects of the standard PIC method.

The flat electron bunch injection is seen in Fig. 3, where the results of the simulations of the ultrashort laser pulse interaction with an underdense plasma target are shown. Here the linearly polarized laser pulse with the electric field along the z direction and with wavelength $\lambda$=0.8 μm has its intensity equal to $I$=10$^{20}$ W/cm$^2$. The pulse duration is 27 fs, and is focused into a spot with a diameter of 16 μm. The laser pulse propagates along the x direction from the right to left in a plasma with density $n_e$=10$^{19}$ cm$^{-3}$. The simulations were performed with the use of "the moving window" technique in a simulation box of size 80x56x56 $\lambda^3$. The mesh sizes in the direction of the laser pulse propagation and in the transverse direction are $\Delta x$= $\lambda$/20 and $\Delta y$=$\Delta z$= $\lambda$/10, respectively, with 8 particles (electrons and protons) per cell. In Fig. 3 where the electron density distribution in the z=0 plane: a); in the y=0 plane: b); and in the $x+ct=315\,\lambda$ plane: c) are presented we distinctly see fast electron bunches injected into the second period of the wake wave. The electron bunch width in the z-direction is approximately two times larger than in the y-direction. The flat electron bunch formation is also seen in Fig. 4, where the density distributions of fast electrons (with $p_x$>200 MeV) in the z=0 plane: a); and in the y=0 plane are presented. We note here that electron oscillations in the perpendicular directions, along the y and z axes, have different frequencies, as distinctly seen in Fig.4.

In the case under consideration the transverse asymmetry of the wake breaking and the electron bunch generation occurs in this case due to the effect of the linear polarization of the laser. As found in Ref. [33] the linearly polarized laser pulse generates axially asymmetric self-focusing channels and a wake with different amplitudes in the directions along and perpendicular to the polarization direction. We note that the wake field accelerated electron bunches with the elliptic form in the transverse direction have been observed in the experiments presented in Ref. [34], where the transverse elongation has been attributed to the effects of linear polarization of the laser pulse-driver.

## 5. Equilibrium and Betatron Oscillations of a Transversally Elongated Electron Bunch with Space Charge

Here we consider the electron bunch equilibrium configuration inside the wake when its transverse size is substantially smaller than its length. Then we describe betratron oscillations of the bunch in the transverse direction. We assume that the longitudinal (along the $x$ axis) scale length of the fast electron bunch is much greater than its scale length in the transverse direction. Below we therefore assume that the wakefield and the electron bunch are homogeneous along the $x$ axis. Such approximation may be valid in the near-axis region of the wake when the injection time is of the order of the electron acceleration time.

Betatron oscillations of electrons moving inside the wake wave occur due to the transverse component of the wake electric field vanishing along the axis and being linearly proportional to the transverse coordinates, $y$ and $z$, in the vicinity of the axis: $\boldsymbol{E}_\perp = E_y \boldsymbol{e}_y + E_z \boldsymbol{e}_z$ with the components, $E_y$ and $E_z$, depending on the specific form of the wake field. The effects of the magnetic field, which is self-generated in the regular wake wave, are substantially weaker than the electric field effects, e. g. see Ref. [17], and we shall neglect them in our model for the sake of simplicity. On the other hand, the pinching by the magnetic field generated by the electric current carried by fast electrons partially compensates the repelling force due to the electron space charge and we incorporate its effects into our description.

We assume that the transverse cross section of the wake wave has an elliptic form with the semi axes equal to $R_W A_{22}$ and $R_W A_{33}$ with $R_W$ being the transverse scale length. The positive electric charge density in the wake is $en_0$. Using the Dirichlet formula for the electric field of a uniformly charged elliptic cylinder (see in Ref. [35] explanations of the Dirichlet formalism for the solution of the Poisson equation in the confocal ellipsoidal coordinates), we write for the transverse electric field originating from the electric charge separation inside the wake:

$$E_y^{WF} \mathbf{e}_y + E_z^{WF} \mathbf{e}_z = \frac{4\pi e n_0}{A_{22} + A_{33}} \left( A_{33} y \mathbf{e}_y + A_{22} z \mathbf{e}_z \right). \tag{9}$$

Within the framework of the test particle approximation, when we can neglect the effects of the electric and magnetic field produced by the fast electron bunch, the relativistic electron motion in the electric field given by Eq. (9) corresponds to the betatron oscillations. It is easy to obtain that the oscillations are performed along the y and z axes with the frequencies

$$\omega_1 = \omega_b \sqrt{\frac{A_{33}}{A_{22} + A_{33}}}, \qquad \omega_2 = \omega_b \sqrt{\frac{A_{22}}{A_{22} + A_{33}}}, \tag{10}$$

which have different values for the oscillations along the y, $\omega_1$, and z axes, $\omega_2$, respectively. Here $\omega_b = \omega_{pe} / \sqrt{\gamma_e}$ with the electron gamma factor, $\gamma_e$, and the Langmuir frequency $\omega_{pe} = \sqrt{4\pi n_0 e^2 / m_e}$. The structure of the mode is given by the relationships:

$$\begin{pmatrix} \delta a_{22} \\ \delta a_{33} \end{pmatrix} = \begin{pmatrix} \sum_{\pm} C_{1,\pm} \exp(\pm i \omega_1 t) \\ \sum_{\pm} C_{2,\pm} \exp(\pm i \omega_2 t) \end{pmatrix}, \tag{11}$$

where $C_{1,\pm}$ and $C_{2,\pm}$ are constants determined by initial conditions. Here and below we assume for the sake of simplicity that the electron gamma factor, $\gamma_e$, does not depend on time. The time dependence of $\gamma_e$ can easily be incorporated into our model as similarly done in Ref. [17], where the betatron oscillations have been studied assuming axial symmetry of the wake and electron bunch.

Now we take into account the space charge effects which generate electric and magnetic fields from the electron bunch. As has been done above, in order to find the electric and magnetic fields $\mathbf{E}_\perp^b$ and $\mathbf{B}_\perp^b$ generated by the elliptic cylindrical electron bunch, we use the Dirichlet formulae. We cast these into the form

$$E_y^b \mathbf{e}_y + E_z^b \mathbf{e}_z = \frac{4eN_b}{r_b^2(a_{22}+a_{33})}\left(a_{33}y\mathbf{e}_y + a_{22}z\mathbf{e}_z\right), \tag{12}$$

and

$$B_y^b \mathbf{e}_y + B_z^b \mathbf{e}_z = \frac{4eN_b v_b}{r_b^2 c(a_{22}+a_{33})}\left(a_{33}z\mathbf{e}_y - a_{22}y\mathbf{e}_z\right). \tag{13}$$

Here $r_b a_{22}$ and $r_b a_{33}$ are the electron bunch semi-axes in the transverse plane, $v_b$ and $N_b$ are the electron velocity along the $x$ axis and the number of electrons per unit length, and $r_b$ is a typical transverse size of the bunch.

The equations of the motion of a fluid element of the electron bunch in the transverse direction are

$$\partial_t n_b + \nabla_\perp (\mathbf{v}_\perp n_b) = 0, \tag{14}$$

$$\partial_t \mathbf{p}_\perp + (\mathbf{v}_\perp \cdot \nabla_t)\mathbf{p}_\perp = e(\mathbf{E}_\perp + \mathbf{v}_\perp \times \mathbf{B}_\perp / c). \tag{15}$$

Here the electron density, $n_b(y,z,t)$, and transverse component of the electron momentum $\mathbf{p}_\perp(y,z,t) = p_y(y,z,t)\mathbf{e}_y + p_z(y,z,t)\mathbf{e}_z$ depend on the coordinates, $y,z$, and time, $t$. The operator $\nabla_\perp$ is given by $\nabla_\perp = \partial_y \mathbf{e}_y + \partial_z \mathbf{e}_z$.

For the electric and magnetic fields linearly dependent on the coordinates, as given by Equations (9,12,13), the equations of the bunch motion admit a self-similar solution, which describes the fluid motion with homogeneous deformation [36]. Within the framework of the homogeneous deformation approximation, the relationship between the Euler, $x_i$, and Lagrange coordinates, $x_i^0$, has the form

$$x_i = a_{ij}(t)x_j^0, \tag{16}$$

where $a_{ij}$ is a deformation matrix with time dependent components. A summation over repeating indices is assumed. Differentiating this relationship with respect to time, we find that the velocity of the electron fluid element is given by $v_i = w_{ij}(t)x_j$ with $w_{ij} = \dot{a}_{ik}a_{kj}^{-1}$. Here $a_{kj}^{-1}$ is the inverse matrix to the matrix $a_{ij}$. A kinematical interpretation of the velocity gradient matrix $w_{ij}$ is provided by analyzing the relative motion of two neighboring fluid particles [37]. The particles we consider are separated by $\delta x_i$. The relative velocity $\delta v_i$ can be written as $\delta v_i = \partial_j v_i \delta x_j = w_{ij}\delta x_j = \Xi_{ij}\delta x_j + \Omega_{ij}\delta x_j$. Here $\partial_i = e_y \partial_y + e_z \partial_z$; the tensors $\Xi_{ij}$ and $\Omega_{ij}$ are determined by $\Xi_{ij} = (\partial_j v_i + \partial_i v_j)/2$ and $\Omega_{ij} = (\partial_j v_i - \partial_i v_j)/2 = -\varepsilon_{ijk}\omega_k/2$ with $\varepsilon_{ijk}$ being the antisymmetric Ricci tensor. The vector $\omega_k$ is the fluid vorticity $\boldsymbol{\omega} = \nabla \times \boldsymbol{v}$. The term $\Xi_{ij}\delta x_j$ describes pure straining motion, while $\Omega_{ij}\delta x_j$ describes rigid body rotation.

In the approximation $|\boldsymbol{p}_\perp| \ll p_x$ the transverse component of the momentum can be written as $\boldsymbol{p}_\perp = m_e \gamma_e (v_y \boldsymbol{e}_y + v_z \boldsymbol{e}_z)$ with the gamma factor, $\gamma_e = (1 - v_b^2/c^2)^{-1/2}$, calculated for the longitudinal energy of fast electrons. In this case we obtain $p_{t,i} = m_e \gamma_e w_{ij} x_j \equiv m_e \gamma_e \dot{a}_{ij} x_j^0$.

Further we shall consider the case of curl free motion, $\nabla_\perp \times \boldsymbol{p}_\perp = 0$, i. e. the matrix $\Omega_{ij}$ vanishes ( $\Omega_{ij} = 0$ ). This corresponds to the diagonal form of the deformation matrix: $a_{ij} = \text{diag}\{1, a_{22}, a_{33}\}$. Assuming the electron density to be homogeneous and substituting the expression $v_i = w_{ij}(t)x_j$ with $w_{ij} = \dot{a}_{ik}a_{kj}^{-1}$ for the electron velocity to the continuity equation Eq. (13), we find that the electron density inside the bunch is given by

$$n_b(t) = n_b(0)\left(\frac{\det a_{ij}(0)}{\det a_{ij}(t)}\right) \tag{17}$$

with $\det a_{ij}$ being the determinant of the matrix $a_{ij}$. In the case under consideration it equals $a_{22}a_{33}$.

In order to illustrate the property of the motion with homogeneous deformation, we consider the simplest example of the dynamics of a pressure-less gas for which the deformation matrix obeys the equation $\ddot{a}_{ij} = 0$ with initial conditions $a_{ij} = \delta_{ij}$ and $\dot{a}_{ij}(0) = w_{ij}(0)$. The solution to this equation is $a_{ij} = \delta_{ij} + w_{ij}(0)t$. The catastrophe corresponds to the situation when the determinant of the matrix $a_{ij}$ vanishes. If the initial matrix of the fluid velocity gradients is diagonal, $w_{ij}(0) = \text{diag}\{1, w_{22}(0), w_{33}(0)\}$, the deformation matrix is equal to $a_{ij}(t) = \text{diag}\{1, 1 + w_{22}(0)t, 1 + w_{33}(0)t\}$. Its determinant is $\det a_{ij}(t) = (1 + w_{22}(0)t)(1 + w_{33}(0)t)$. A singularity occurs when either $t = -1/w_{22}(0)$ or $t = -1/w_{33}(0)$. The singularity occurs as a line in 3D space for $w_{22}(0)$ and $w_{33}(0)$ being equal and both negative, and singularity appears as a surface in 3D space when just one value among $w_{22}(0)$ and $w_{33}(0)$ is negative. The generic case corresponds to the situation when just one value among $w_{22}(0)$ and $w_{33}(0)$ is negative. This means that in the generic case the singularity develops as a surface. We note that such type of singularity on a surface has been studied in detail in applications for the nonlinear dynamics of gravitational instability [38] and in the theory of magnetic field line reconnection in high conductivity plasmas [39].

Since the number of electrons per unit length of the elliptical cylinder with the semi-axes $a = r_b a_{22}$ and $b = r_b a_{33}$ is equal to $N_b = n_b \pi r_b^2 a_{22} a_{33}$ we can rewrite Eq. (17) as $n_b = N_b / \pi r_b^2 a_{22} a_{33}$. From Eqs. (9,12,13,15,17) we obtain for the matrix $a_{ij}$ components a system of ordinary differential equations

$$\ddot{a}_{22} = -\frac{4\pi e^2}{m_e \gamma_e} \left[ \frac{A_{33} n_0 a_{22}}{(A_{22} + A_{33})} - \frac{N_b}{\pi r_b^2 \gamma_e^2 (a_{22} + a_{33})} \right], \tag{18}$$

$$\ddot{a}_{33} = -\frac{4\pi e^2}{m_e \gamma_e} \left[ \frac{A_{22} n_0 a_{33}}{(A_{22} + A_{33})} - \frac{N_b}{\pi r_b^2 \gamma_e^2 (a_{22} + a_{33})} \right]. \tag{19}$$

We see similarity between these equations and the equations for the charged particle beam dynamics in the high energy accelerators, which are obtained within the framework of the Kapchinskij – Vladimirskij approximation [40] (e. g. see Refs. [41,42]).

Using notations, $a = r_b a_{22}$, $b = r_b a_{33}$, $K_2 = 4\pi n_0 e^2 A_{33} / m_e c^2 \gamma_e (A_{22} + A_{33})$, $K_3 = 4\pi n_0 e^2 A_{22} / m_e c^2 \gamma_e (A_{22} + A_{33})$, and incorporating the transverse emittance effects, we rewrite equations (18, 19) as

$$a'' + K_2 a = \frac{\varepsilon_2^2}{a^3} + \frac{\xi}{a+b}, \qquad (20)$$

$$b'' + K_3 b = \frac{\varepsilon_3^2}{b^3} + \frac{\xi}{a+b}. \qquad (21)$$

Here $\varepsilon_2$ and $\varepsilon_3$ are the transverse emittance values in the y and z directions, $\xi = N_b / \pi n_0 r_b^2 \gamma_e^2$ is a dimensionless space charge parameter, and the primes denote differentiation with respect to the variable $s = ct$. Properties of this equation system are discussed in detail in Ref. [40].

Equations (20,21) can be cast into the Hamiltonian form with the Hamiltonian depending on the canonical coordinates $a$ and $b$, and on the canonical momenta $\pi_2$ and $\pi_3$. It reads

$$\mathcal{H}(\pi_2, \pi_3, a, b) = \frac{1}{2}\left( \pi_2^2 + \pi_3^2 + K_2 a^2 + K_3 b^2 + \frac{\varepsilon_2^2}{a^2} + \frac{\varepsilon_3^2}{b^2} \right) - \xi \ln(a+b) \qquad (22)$$

from which we may conclude that for $a > 0$, $b > 0$ the bunch performs nonlinear oscillations around the equilibrium. In Fig. 5 we plot iso-contours of the potential function,

$$\Pi(a,b) = \frac{1}{2}\left( K_2 a^2 + K_3 b^2 + \frac{\varepsilon_2^2}{a^2} + \frac{\varepsilon_3^2}{b^2} \right) - \xi \ln(a+b), \qquad (23)$$

for $K_2 = 2.5, K_3 = 5, \varepsilon_2 = 1, \varepsilon_3 = 1, \xi = 1$ in frame a), and for $K_2 = 1, K_3 = 2.25, \varepsilon_2 = 1, \varepsilon_3 = 1, \xi = 5$ in frame b). From the form of the potential function, $\Pi(a,b)$, we can see that in general case the frequencies of the oscillations along the y- and z-directions are different and depend on the oscillation amplitudes. If the emittances, $\varepsilon_2$ and $\varepsilon_3$, which are determined by the injection mechanism, vanish, as in Fig. 5 b), the iso-contours of the function $\Pi(a,b)$ can intersect the axis *a=0* or *b=0*. This corresponds to the case when in nonlinear oscillations one semi-axis of the bunch becomes equal to zero and formally the bunch aspect ratio tends to infinity. We note that in the axially symmetric configuration the space charge effect prevents the bunch radius from vanishing. For a flat electron beam the space charge effects are not strong enough and the bunch demagnification in one of the directions becomes possible.

The transverse equilibrium of the electron bunch in the wake corresponds to the local minimum of the function $\Pi(a,b)$ given by Eq. (23). It is given by a static solution of Eqs. (20,21) for which the terms on the right hand sides vanish. We consider the case when the emittances vanish. Solving these algebraic equations assuming $\varepsilon_2 = 0$ and $\varepsilon_3 = 0$, we obtain for the equilibrium

$$a^{eq} = \sqrt{\xi \frac{K_2}{K_3}}, \qquad b^{eq} = \sqrt{\xi \frac{K_3}{K_2}}. \tag{24}$$

We point out that in equilibrium the electron bunch has an elliptic cross section with the aspect ratio $a^{eq}/b^{eq}$ equal to the aspect ratio of the wake: $a^{eq}/b^{eq} = A_{22}/A_{33}$. The electron density inside the bunch in the equilibrium is equal to

$$n_b^{eq} = \frac{N_b}{\pi r_b^2 ab} = n_0 \gamma_e^2. \tag{25}$$

It is by a factor $\gamma_e^2$ greater than the ion density in the plasma. The characteristic transverse size of the bunch, $r_b$, can be found to be

$$r_b = \sqrt{\frac{N_b}{\pi n_0 \gamma_e^2}}, \tag{26}$$

i. e. the dimensionless space charge parameter, $\xi = N_b / \pi n_0 r_b^2 \gamma_e^2$, for the equilibrium configuration is equal to unity, $\xi = 1$.

Using the definition (26) for the transverse size of the bunch, $r_b$, and linearizing equations (20,21) in the vicinity of the equilibrium solution (24) for $\varepsilon_2 = 0$ and $\varepsilon_3 = 0$, $a^{eq} = \sqrt{A_{22}/A_{33}}$, $b^{eq} = \sqrt{A_{33}/A_{22}}$, i.e. representing the functions $a$ and $b$ in the form $a = a^{eq} + \delta a$, $b = b^{eq} + \delta b$ with $\delta a \ll a$, $\delta b \ll b$, we obtain

$$\delta \ddot{a} = -\omega_b^2 \left\{ \left[ \frac{A_{33}}{(A_{22}+A_{33})} + \frac{A_{22}A_{33}}{(A_{22}+A_{33})^2} \right] \delta a + \frac{A_{22}A_{33}}{(A_{22}+A_{33})^2} \delta b \right\}, \tag{27}$$

$$\delta \ddot{b} = -\omega_b^2 \left\{ \frac{A_{22}A_{33}}{(A_{22}+A_{33})^2} \delta a + \left[ \frac{A_{22}}{(A_{22}+A_{33})} + \frac{A_{22}A_{33}}{(A_{22}+A_{33})^2} \right] \delta b \right\}. \tag{28}$$

These equations describe oscillations with frequencies

$$\omega_1^b = \omega_b, \qquad \omega_2^b = \omega_b \sqrt{\frac{2A_{22}A_{33}}{A_{22}+A_{33}}}. \tag{29}$$

As we see, in the case $A_{22} \ll A_{33}$ the frequency $\omega_2^b$ is much lower than $\omega_1^b$. We also see that the frequency values in the case when the space charge effect is taken into account, Eq. (29), are different from the frequencies, Eq. (10), obtained within the framework of the test particle approximation.

The structure of the mode is described by the relationships:

$$\begin{pmatrix} \delta a \\ \delta b \end{pmatrix} = \frac{1}{\left(A_{22}^2 + A_{33}^2\right)} \begin{pmatrix} A_{33} & A_{22} \\ -A_{22} & A_{33} \end{pmatrix} \begin{pmatrix} \sum_{\pm} C_{1,\pm} \exp(\pm i\omega_1^b t) \\ \sum_{\pm} C_{2,\pm} \exp(\pm i\omega_2^b t) \end{pmatrix}, \qquad (30)$$

where $C_{1,\pm}$ and $C_{2,\pm}$ are constants given by initial conditions. This expression corresponds to the squinted ellipse form of the potential energy iso-contours in the vicinity of the bunch equilibrium position presented in Fig. 5 b).

## 6. Discussion and Conclusion

In conclusion, a method is suggested for electron injection via the transverse wake wave breaking, when the electron trajectory self-intersection leads to the formation of an electron bunch elongated in the transverse direction. In this scheme we take advantage of the laser pulse focussed into an elongated spot. In turn, this results in a wakefield generation that is localized in the axially non-symmetric region with the components of the transverse electric field not equal to each other. With the aid of catastrophe theory we demonstrate that a structurally stable regime of transverse wake breaking leads to the transversally elongated electron bunch generation. Three-dimensional particle-in-cell simulations of the laser pulse interaction with an underdense plasma show that electrons, injected via the transverse wake wave breaking, form a bunch with an aspect ratio larger than unity. Electrons accelerated by the wake perform betatron oscillations with different amplitudes and frequencies along the two transverse coordinates. An exact analytical solution of the electron hydrodynamics equations demonstrates the space charge effects, which modify the electron bunch equilibrium and the frequencies and structure of the mode of betatron oscillations.

For a typical total number of electrons in the bunch accelerated by the wake equal to $N_{tot} = 10^{10}$, which corresponds to the charge 1.6 $nC$, and a bunch length of 10 $\mu m$, the electron number per unit length is $N_b = 10^{13} cm^{-1}$. If the plasma density and electron gamma factor are $n_0 = 10^{19} cm^{-3}$ and $\gamma_e = 500$, expression (26) yields that the bunch size in the transverse direction is approximately equal to $r_b \approx 0.01\ \mu m$. Betatron frequencies, Eq. (29), in this case are $\omega_1^b \approx 10^{14}\ s^{-1}$ and $\omega_2^b \approx 1.4 \times 10^{13}\ s^{-1}$, if we assume that the aspect ratio, $A_{22}/A_{33}$, equals 100.

One of the most important applications of laser produced relativistic electrons is to generate positron bunches for their injection into conventional accelerators. To obtain polarized positron beams an additional circularly polarized laser pulse (counterpropagating with the electron bunch) may be used to generate longitudinally polarized gamma ray photons, which then collide with a thin film target [43].

The emittance of laser accelerated ions can also be manipulated by changing the structure of either the target irradiated by the laser pulse or the form of the focusing system. In the first case we refer to the double layer target proposed in Ref. [44] in order to produce beams with controlled quality, and studied in detail via computer simulations [45] and experiments [46]. Ref. [44] proposed to use two-layer targets in which the first layer consists of heavy multicharged ions and the second layer (thin and narrow in the transverse direction) consists of light ions (e.g. protons). If the thin proton layer is elongated in one direction it will result in the generation of a flat proton beam. A more complex form of the proton layer may be used in order to provide a uniform irradiation of the target, which is required in the applications of the laser accelerated ions for hadron therapy. The second case uses an ion focusing technique corresponding to a thin hollow cylindrical shell irradiated by a femtosecond high-power laser pulse when the ion bunch flies through it. As demonstrated in the experiments presented in Ref. [47], this technique allows one to simultaneously focus the proton beam and to cut it onto quasi-monoenergetic beamlets. The use of an elliptical cylinder shell provides a way for the transverse emittance manipulation. In addition, a phase rotator, which also produces quasi-monoenergetic ion beamlets [48], when its transverse electric field is made anisotropic, can produce flat ion beams.

## Acknowledgments

This work is supported by the Ministry of Education, Science, Sports and Culture, Grand-in-Aid for Specially Promoted Research No. 15002013. One of the authors (M. K.) is supported by the Ministry of Education, Science, Sports and Culture, Grant-in-Aid for Young Scientists (B), No. 17740272, 2005.

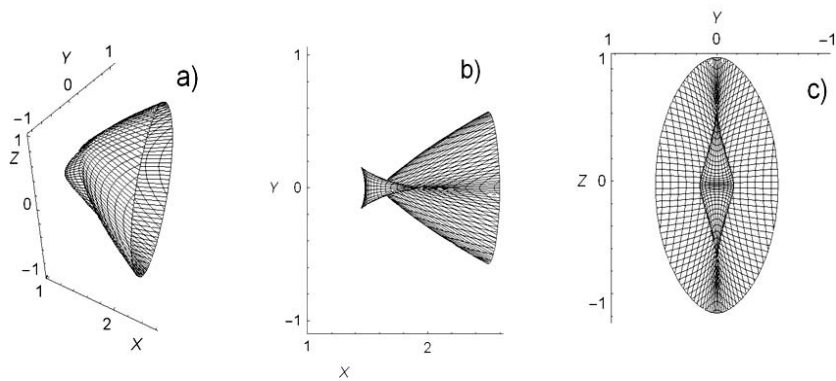

Figure 1. The constant phase surface for $R_y = 1$, $R_z = 1.666$ and $\varsigma = 1.5$, in the $x, y, z$ space: a); and its projections on the $(x, y)$: b); and $(y, z)$: c) planes.

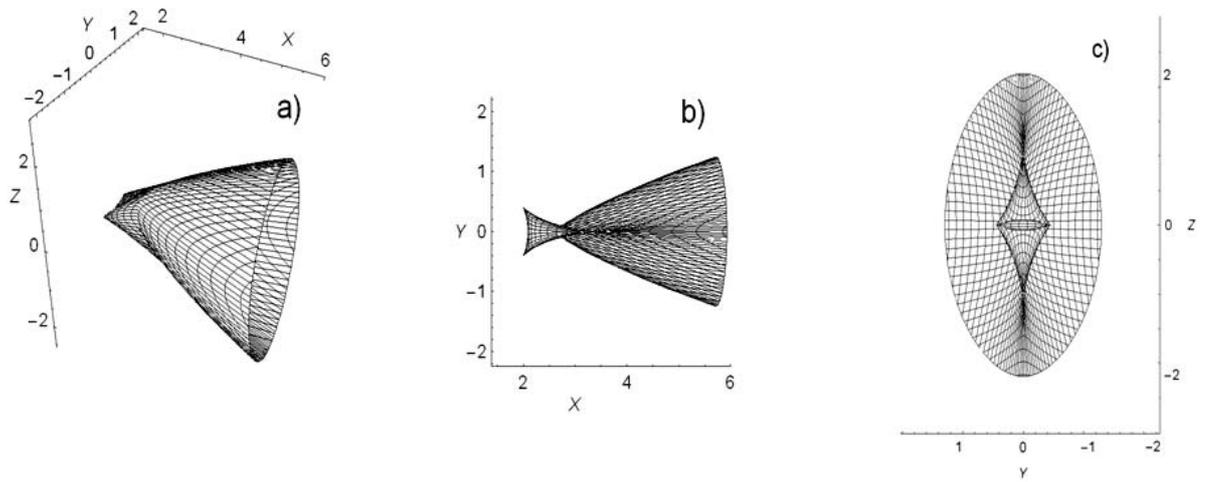

Figure 2. The constant phase surface for $R_y = 1$, $R_z = 1.666$ and $\varsigma = 2$, in the $x, y, z$ space: a); and its projections on the $(x, y)$: b); and $(y, z)$: c) planes.

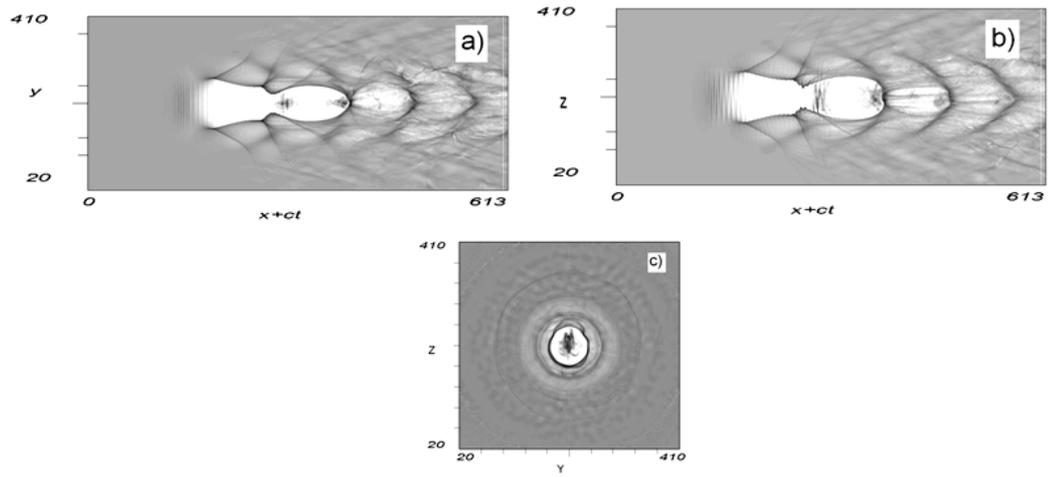

Figure 3. Results of the 3D PIC simulations: the electron density distribution in the z=0 plane: a); in the y=0 plane: b); and in the $x+ct=315\lambda$ plane: c).

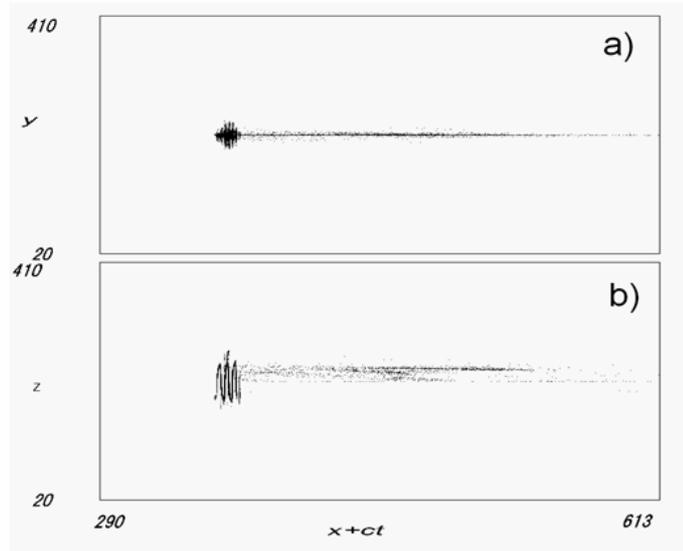

Figure 4. Fast electron (with $p_x>200$ MeV) density distribution in the z=0 plane: a); and in the y=0 plane.

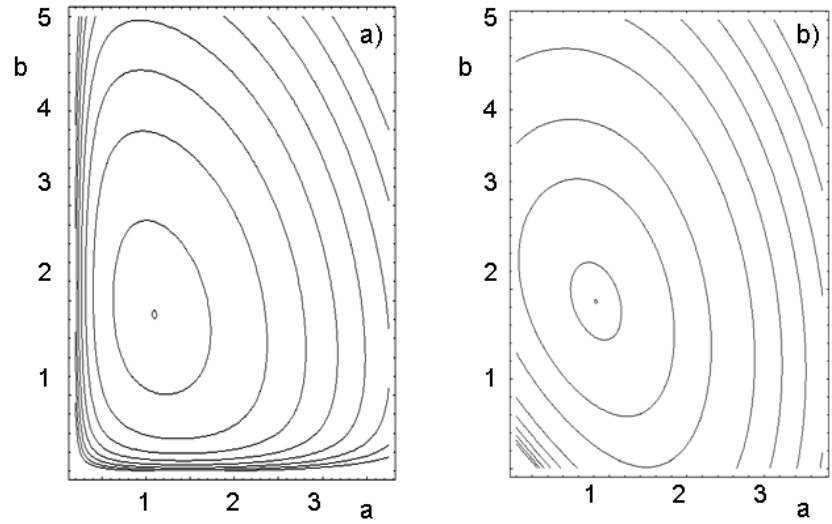

Figure 5. Iso-contours of the potential function $\Pi(a,b)$ for a) $K_2 = 2.5, K_3 = 5, \varepsilon_2 = 1, \varepsilon_3 = 1, \xi = 1$, and b) for $K_2 = 1, K_3 = 2.25, \varepsilon_2 = 1, \varepsilon_3 = 1, \xi = 5$.